\documentclass[letter,twocolumn]{jpsj2} 
\title
{
Singlet Ground State and Spin Gap in ${\mib S}$=1/2 Kagom\'{e} Antiferromagnet Rb$_2$Cu$_3$SnF$_{12}$
}

\author
{ 
Katsuhiro \textsc{Morita}, Midori \textsc{Yano}, Toshio \textsc{Ono}, Hidekazu \textsc{Tanaka}\thanks{E-mail address: tanaka@lee.phys.titech.ac.jp}, Kotaro \textsc{Fujii}$^{1}$, Hidehiro \textsc{Uekusa}$^{1}$,\\
Yasuo \textsc{Narumi}$^{2}$ and Koichi \textsc{Kindo}$^{2}$
}

\inst
{
Department of Physics, Tokyo Institute of Technology, Oh-okayama, Meguro-ku, Tokyo 152-8551\\
$^{1}$Department of Chemistry, Tokyo Institute of Technology, Oh-okayama, Meguro-ku, Tokyo 152-8551\\
$^{2}$Institute for Solid State Physics, The University of Tokyo, Kashiwanoha, Kashiwa, Chiba 277-8581
}

\recdate
{
\hspace{7cm}
}

\abst{We investigated the crystal structure of Rb$_2$Cu$_3$SnF$_{12}$ and its magnetic properties using single crystals. This compound is composed of Kagom\'{e} layers of corner-sharing CuF$_{6}$ octahedra with a $2a{\times}2a$ enlarged cell as compared with the proper Kagom\'{e} layer. Rb$_2$Cu$_3$SnF$_{12}$ is magnetically described as an $S$=1/2 modified Kagom\'{e} antiferromagnet with four kinds of neighboring exchange interaction. From magnetic susceptibility and high-field magnetization measurements, it was found that the ground state is a disordered singlet with the spin gap, as predicted from a recent theory. Exact diagonalization for a 12-site Kagom\'{e} cluster was performed to analyze the magnetic susceptibility, and individual exchange interactions were evaluated.}

\kword{Rb$_2$Cu$_3$SnF$_{12}$, Kagom\'{e} lattice, frustration, singlet ground state, spin gap}

\begin{document}
\maketitle

Antiferromagnets on highly frustrated lattices produce a rich variety of physics \cite{Harrison,ML}. In particular, a two-dimensional Heisenberg Kagom\'{e} antiferromagnet (2D HKAF) is of great interest from the viewpoint of the interplay of the frustration and quantum effects. There are many theoretical studies on the 2D HKAF. The spin wave theory for a large spin value predicted an ordered ground state with the so-called $\sqrt{3}\,{\times}\,\sqrt{3}$ structure, which is selected by quantum fluctuation from infinite classical ground states \cite{Harris,Chubukov}, whereas for a small spin value, a disordered ground state was observed by various approaches \cite{Zeng,Sachdev,Chalker,Elstner,Nakamura}. Recent careful analyses and numerical calculations for an $S$=1/2 case demonstrated that the ground state is a spin liquid state composed of singlet dimers only, and that the ground state is gapped for triplet excitations, but gapless for singlet excitations.\cite{Lecheminant,Waldtmann,Mambrini} Consequently, magnetic susceptibility has a rounded maximum at $T{\sim}(1/6)J/k_{\rm B}$ and decreases exponentially toward zero with decreasing temperature, while specific heat exhibits a power law behavior at low temperatures \cite{Elstner,Misguich}. Specific heat also shows an additional structure, peak or shoulder at low temperatures after exhibiting a broad maximum at $T{\sim}(2/3)J/k_{\rm B}$. 

The experimental studies of the $S$=1/2 HKAF have been limited, and the above-mentioned intriguing predictions have not been verified experimentally. The cupric compounds Cu$_3$V$_2$O$_7$(OH)$_2$$\cdot$2H$_2$O \cite{Hiroi}, $\beta$-Cu$_3$V$_2$O$_8$, \cite{Rogado} and [Cu$_3$(titmb)$_2$(CH$_3$CO$_2$)$_6$]$\cdot$H$_2$O\,\cite{Honda} are known to have a Kagom\'{e} or closely related lattice. The Kagom\'{e} net is distorted into an orthorhombic form in Cu$_3$V$_2$O$_7$(OH)$_2$$\cdot$2H$_2$O and a staircase in $\beta$-Cu$_3$V$_2$O$_8$, so that the exchange network is anisotropic. For Cu$_3$V$_2$O$_7$(OH)$_2$$\cdot$2H$_2$O, no magnetic ordering was observed down to 50 mK \cite{Fukaya}, but the magnetic susceptibility shows no tendency to decrease toward zero.\cite{Hiroi} For $\beta$-Cu$_3$V$_2$O$_8$, magnetic ordering occurs at 29 K \cite{Rogado}. In [Cu$_3$(titmb)$_2$(CH$_3$CO$_2$)$_6$]$\cdot$H$_2$O, the nearest neighbor exchange interaction is ferromagnetic \cite{Narumi}. Recently, the herbertsmithite ZnCu$_3$(OH)$_6$Cl$_2$ with the proper Kagom\'{e} lattice has attracted considerable attention \cite{Shores,Mendels,Helton,Bert,Lee,Rigol,Misguich2}. Although no magnetic ordering occurs down to 50 mK,\cite{Mendels} magnetic susceptibility exhibits a rapid increase at low temperatures.\cite{Helton,Bert} This behavior was ascribed to a large number of idle spins ($4{\sim}10$\,\%) produced by intersite mixing between Cu$^{2+}$ and Zn$^{2+}$\,\cite{Bert,Lee,Misguich2} and/or the Dzyaloshinsky-Moriya interaction.\cite{Rigol} The singlet ground state has not been observed in these systems. The search for new 2D HKAFs with $S$=1/2 continues.

Cs$_2$Cu$_3$ZrF$_{12}$ and Cs$_2$Cu$_3$HfF$_{12}$ have the proper Kagom\'{e} layer at room temperature\,\cite{Mueller} and are promising $S$=1/2 HKAFs.\cite{Yamabe} Unfortunately, these systems undergo structural phase transitions at $T_{\rm t}=220$ and 170 K, respectively, and also magnetic phase transitions at $T_{\rm N}\simeq 24$ K.\cite{Yamabe} However, the magnetic susceptibilities observed at $T > T_{\rm t}$ can be perfectly described using theoretical results for an $S=1/2$ HKAF with large exchange interactions $J/k_{\rm B}\sim 250$ K.\cite{Morita} 

In the present work, we synthesized the new hexagonal compound Rb$_2$Cu$_3$SnF$_{12}$ with a similar crystal structure as Cs$_2$Cu$_3$ZrF$_{12}$ and performed magnetic susceptibility and high-field magnetization measurements using single crystals. As shown below, we found that the ground state is a disordered singlet with a finite gap for magnetic excitations. 

Rb$_2$Cu$_3$SnF$_{12}$ crystals were synthesized via the chemical reaction $\mathrm{2RbF + 3CuF_2 + SnF_4}$ $\rightarrow$ $\mathrm{Rb_2Cu_3SnF_{12}}$. 
$\mathrm{RbF}$, $\mathrm{CuF_2}$, and $\mathrm{SnF_4}$ were dehydrated by heating in vacuum at $60{\sim}100^{\circ}$C for three days. The materials were sealed in a platinum tube in the ratio of $3:3:2$. Single crystals were grown from the melt. The temperature of the furnace was lowered from 800 to 550$^{\circ}$C for four days. Transparent colorless crystals with a typical size of $5{\times}5{\times}0.5$ mm$^3$ were obtained. 
 
Since the crystal structure of Rb$_2$Cu$_3$SnF$_{12}$ has not been reported to date, we performed a structural analysis at room temperature using a Bruker SMART-1000 three-circle diffractometer equipped with a CCD area detector. Monochromatic Mo-K$\alpha$ radiation was used as an X-ray source. Data integration and global-cell refinements were performed using data in the range of $1.96^\circ<{\theta}< 27.64^\circ$, and multi-scan empirical absorption correction was also performed. The total number of reflections observed was 15268, and 1765 reflections were found to be independent and 1688 reflections were determined to satisfy the criterion $I > 2{\sigma}(I)$. Structural parameters were refined by the full-matrix least-squares method using SHELXL-97 software. The crystal showed a merohedral twin structure. The twin law is $(0,1,0\,/\,1,0,0\,/\,0,0,-1)$ and the occupancy of the twin component was refined to 0.393(1). The final $R$ indices obtained were $R$=0.020 and $wR$=0.056. 
\begin{table}
\caption{Atomic coordinates ($\times 10^4$), equivalent isotropic displacement parameters ($\rm{\AA}^2{\times}10^3$), and site occupancies.}
\label{table:2}
\begin{tabular}{lccccc}\hline
Atom  &  $x$  & $y$  & $z$ & $U_{\rm eq}$ & Occ.  \\ \hline
Rb(1) & 6667 & 3333 & $-$647(1) & 37(1) & 1   \\
Rb(2) & 3358(1) & 1650(1) & 641(1) & 35(1) & 1  \\
Cu(1) & 5946(1) & 1665(1) & 1682(1) & 18(1) & 1  \\
Cu(2) & 3516(1) & $-$723(1) & 1769(1) & 16(1) & 1  \\
Sn(1) & 5000 & 0 & 0 & 16(1) & 1  \\
Sn(2) & 6667 & 3333 & 3333 & 15(1) & 1  \\
F(1) & 6495(2) & 651(2) & 1700(1)  & 23(1) & 1  \\
F(2) & 7409(2) & 2773(2) & 1439(1) & 30(1) & 1  \\
F(3) & 4581(2) & 693(2) & 2105(1) & 30(1) & 1   \\ 
F(4) & 7853(2) & 4443(2) & 2780(1) & 32(1) & 1   \\ 
F(5) & 4464(3) & 872(3) & $-$460(1) & 41(1) & 1   \\ 
F(6A) & 4501(4) & 310(4) & 835(2) & 40(1) & 0.773(8)   \\ 
F(7A) & 6389(3) & 1378(3) & 95(3) & 49(1) &  0.773(8)  \\ 
F(6B) & 3861(11) & $-$250(11) & 629(6) & 29(3) & 0.227(8)   \\ 
F(7B) & 6066(10) & 1320(10) & 470(6) & 23(3) & 0.227(8)   \\  
F(8) & 2465(2) & $-$2156(2) & 1462(1) & 26(1) & 1   \\
\hline
\end{tabular}
\end{table}

\begin{figure}[t]
	\begin{center}
		\includegraphics[width=8.0cm]{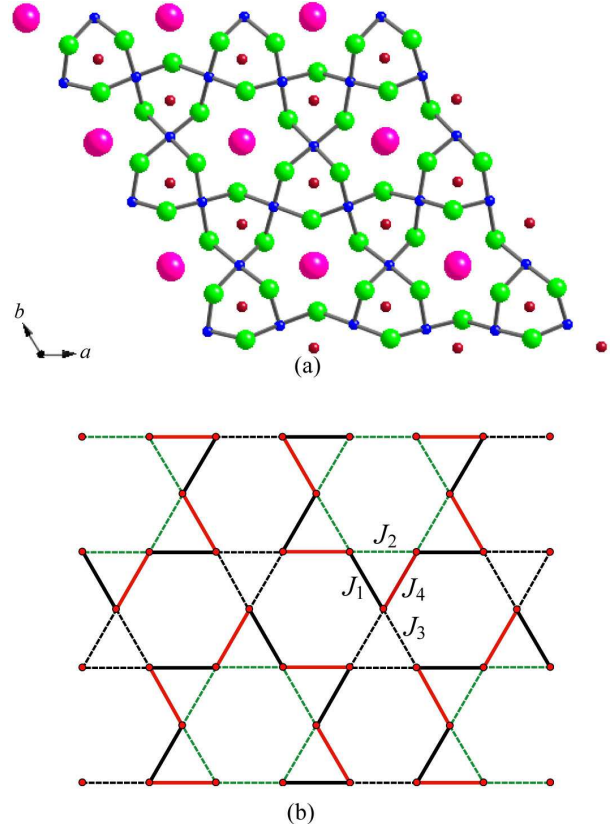}
	\end{center}
	\caption{(a) Crystal structure of Rb$_2$Cu$_{3}$SnF$_{12}$ viewed along the $c$-axis, where F(4), F(5), F(6), and F(7) are omitted. Gray bonds denote exchange pathways. Large red, middle green, and small blue and brown circles are Rb$^+$, F$^-$, Cu$^{2+}$, and Sn$^{4+}$ ions, respectively. Dashed lines denote the unit cell. (b) Exchange network in Kagom\'{e} layer.}
	 \label{fig:StrucEx}
 \end{figure}
 
The structure of Rb$_2$Cu$_3$SnF$_{12}$ is hexagonal, $R{\bar 3}$, with cell dimensions of $a$=13.917(2)\,\AA\ and $c$=20.356(3)\,\AA, and $Z$=12. Atomic coordinates, equivalent isotropic displacement parameters, and site occupancies are shown in Table \ref{table:2}. Disorder was observed for the positions of F(6) and F(7). Site occupancies for F(6A) and F(7A) are 0.773(8) and those for F(6B) and F(7B) are 0.227(8). The crystal structure viewed along the $c$-axis is illustrated in Fig. \ref{fig:StrucEx}(a), where F(4), F(5), F(6), and F(7) located outside the Kagom\'{e} layer are omitted, so that magnetic Cu$^{2+}$ ions and exchange pathways (gray bonds) are visible. The structure of Rb$_2$Cu$_3$SnF$_{12}$ is closely related to the Cs$_2$Cu$_3$ZrF$_{12}$ structure.\cite{Mueller} The chemical unit cell is described by enlarging that of Cs$_2$Cu$_3$ZrF$_{12}$ to $2a$, $2a$, $c$.
CuF$_6$ octahedra are linked in the $c$-plane sharing corners. Magnetic Cu$^{2+}$ ions with $S$=1/2 located at the center of the octahedra form a Kagom\'{e} lattice in the $c$-plane. CuF$_6$ octahedra are elongated along the principal axes that are approximately parallel to the $c$-axis, so that the hole orbitals $d(x^2-y^2)$ of Cu$^{2+}$ spread in the Kagom\'{e} layer. Figure \ref{fig:StrucEx}(b) shows the exchange network in the Kagom\'{e} layer. There are four kinds of nearest-neighbor exchange interaction. Since the bond angle ${\alpha}$ of the exchange pathway Cu$^{2+}$$-$F$^{-}$$-$Cu$^{2+}$ is ${\alpha}$=$123.9{\sim}138.4^{\circ}$, the exchange interactions should be antiferromagnetic and strong. Ferromagnetic superexchange occurs only when ${\alpha}$ is close to $90^{\circ}$. The exchange interactions $J_1{\sim}J_4$ are labeled in decreasing order of ${\alpha}$. Disorder in F(6) and F(7) may slightly affect the exchange interactions between Cu$^{2+}$ ions, because they are outside the Kagom\'{e} layer and do not take part in the exchange processes. The interlayer exchange interaction $J'$ should be much smaller than $J$, because magnetic Cu$^{2+}$ layers are sufficiently separated by nonmagnetic Rb$^+$, Sn$^{4+}$, and F$^-$ layers. Thus, Rb$_2$Cu$_3$SnF$_{12}$ is expected to be a 2D $S$=1/2 HKAF.

Magnetic susceptibilities of Rb$_2$Cu$_{3}$SnF$_{12}$ were measured in the temperature range 1.8$-$400 K using a SQUID magnetometer (Quantum Design MPMS XL). High-field magnetization measurement was performed using an induction method with a multilayer pulse magnet at the Institute for Solid State Physics, The University of Tokyo. Magnetic fields were applied parallel and perpendicular to the $c$-axis in both experiments.

\begin{figure}
\begin{center}
\includegraphics[width=8.0cm]{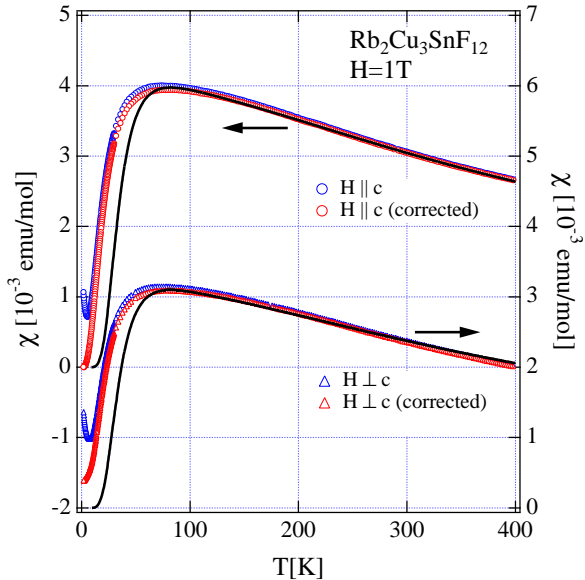}
\end{center}
\caption{Temperature dependence of magnetic susceptibilities of Rb$_2$Cu$_3$SnF$_{12}$. Blue symbols denote raw data. The susceptibilities corrected for the Curie-Weiss term due to the impurity phase are plotted using red symbols. Solid lines denote the results obtained by exact diagonalization for a 12-site Kagom\'{e} cluster with the exchange interactions and $g$-factors shown in the text.}
\label{fig:SusEx}
\end{figure}

Figure \ref{fig:SusEx} shows the temperature dependence of magnetic susceptibilities of Rb$_2$Cu$_3$SnF$_{12}$ measured at $H$=1 T. It is noted that one molar Rb$_2$Cu$_3$SnF$_{12}$ contains three molar Cu$^{2+}$ ions. The obtained susceptibility data were corrected for the diamagnetism ${\chi}_{\rm dia}$ of core electrons and for Van Vleck paramagnetism. The diamagnetic susceptibilities of individual ions were taken from the literature \cite{Selwood}. Van Vleck paramagnetic susceptibility was calculated using
$
{\chi}_{\rm VV}^{\mu}=-(N{\mu}_{\rm B}^2/{\lambda}){\Delta g}_{\mu}=3.14\times 10^{-4}{\Delta g}_{\mu}\ {\rm emu/Cu^{2+}\,mol},
$
where ${\lambda}=-829$ cm$^{-1}$ is the spin-orbit coupling coefficient of Cu$^{2+}$ and ${\Delta g}_{\mu}=g_{\mu}-2$ is the anisotropy of the $g$-factor. The $g$-factors used are $g_{\parallel}$=2.44 and $g_{\perp}$=2.15, which were determined self-consistently by fitting the result of exact diagonalization to the experimental susceptibilities shown in Fig. \ref{fig:SusEx}. These $g$-factors are consistent with those parallel and perpendicular to the elongated axes of CuF$_6$ octahedra in K$_2$CuF$_4$ and Rb$_2$CuF$_4$.\cite{Sasaki}

With decreasing temperature, the magnetic susceptibilities of Rb$_2$Cu$_3$SnF$_{12}$ exhibit rounded maxima at $T_{\rm max}{\sim}70$ K and decrease rapidly. Small upturn is observed below 7 K, which should be ascribed to impurities on crystal surfaces. No magnetic ordering is observed. This result indicates clearly that the ground state is a disordered singlet with a spin gap, as predicted from a recent theory on a 2D $S$=1/2 HKAF. Low-temperature susceptibilities are expressed as
\begin{eqnarray}
{\chi}=C/(T-{\Theta})+A\,{\exp}\left(-{\Delta}/k_{\rm B}T\right)+C_0,
\end{eqnarray} 
where the first term is the Curie-Weiss term due to impurities, the second term, the asymptotic low-temperature susceptibility for a 2D spin gap system,\cite{Stone} and the last term, the constant term arising from the triplet component mixed in the ground state. Applying eq. (1) to  the raw susceptibility data for $T$\,$<$\,10 K, we evaluated the impurity contribution (${\simeq}0.3$\,\%). Red symbols in Fig. \ref{fig:SusEx} denote the intrinsic susceptibilities corrected for impurities. The susceptibility for $H{\parallel}c$ is almost zero for $T{\rightarrow}0$, whereas that for $H{\perp}c$ is finite. In Rb$_2$Cu$_3$SnF$_{12}$, elongated axes of CuF$_6$ octahedra incline alternately in the Kagom\'{e} layer, inducing a staggered field when an external field is applied. Since there is no inversion center in the middle of two neighboring magnetic ions in the Kagom\'{e} layer, the Dzyaloshinsky-Moriya (DM) interaction is allowed. The Zeeman interaction due to the staggered field and the DM interaction can have finite matrix elements between the singlet ground state and the excited triplet state, because they are antisymmetric with respect to the interchange of the interacting spins. Thus, we infer that the ground state has a small amount of triplet component through these antisymmetric interactions when subjected to the external field parallel to the Kagom\'{e} layer. This gives rise to the finite susceptibility at $T$=0. The magnitude of the spin gap is estimated to be ${\Delta}/k_{\rm B}=21(1)$\,K by fitting eq. (1) to the raw susceptibility data for $H{\parallel}c$.
\begin{figure}
\begin{center}
\includegraphics[width=8.0cm]{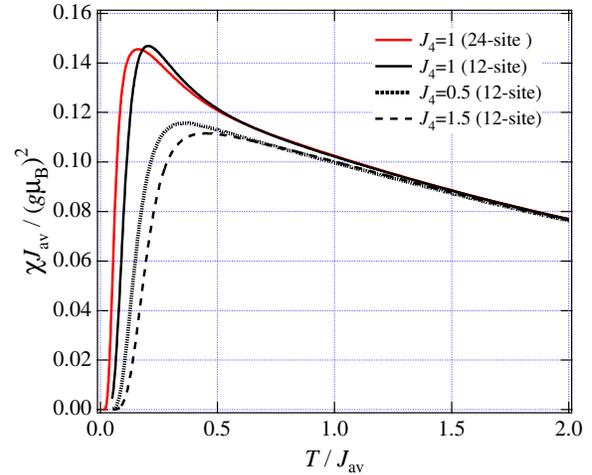}
\end{center}
\caption{Temperature dependence of magnetic susceptibilities per site for $J_4$=0.5, 1, and 1.5 obtained by exact diagonalizations for 12-site Kagom\'{e} cluster, where we set $J_1$=$J_2$=$J_3$=1. Temperature is scaled using the average of the exchange interactions $J_{\rm av}$. The red solid line indicates the result for the uniform KHAF obtained by exact diagonalization for the 24-site Kagom\'{e} cluster.\cite{Misguich2}}
\label{fig:SusTh}
\end{figure}

For $T$\,$<$\,150 K, the magnetic susceptibility of the present system does not agree with the theoretical result for the $S$=1/2 uniform HKAF shown by red solid line in Fig. \ref{fig:SusTh}.\cite{Misguich2} The theoretical susceptibility exhibits a sharp rounded maximum at $T_{\rm max}{\sim}(1/6)J/k_{\rm B}$, whereas the experimental susceptibility exhibits a broad maximum. This is because four nearest-neighbor exchange interactions are different. As shown in Fig. \ref{fig:StrucEx}(b), one chemical unit cell in a Kagom\'{e} layer contains 12 spins. Then, we carried out the exact diagonalization for the 12-site Kagom\'{e} cluster under the periodic boundary condition. First, we calculated the uniform case, $J_1$=$J_2$=$J_3$=$J_4$=1, to check the validity of our calculation. The obtained result is shown by a black solid line in Fig. \ref{fig:SusTh}, where temperature and susceptibility are scaled using the average of exchange interactions $J_{\rm av}$\,=\,$(1/4)\sum_i J_i$. The result is the same as that obtained by Elstner and Young,\cite{Elstner} and close to the result for the 24-site Kagom\'{e} cluster obtained by Misguich and Sindzingre.\cite{Misguich2} Dashed and dotted lines in Fig. \ref{fig:SusTh} denote the results for $J_4$=1.5 and 0.5, respectively, where we set $J_1$=$J_2$=$J_3$=1. Whether for $J_4>1$ or $J_4<1$, the maximum susceptibility ${\chi}_{\rm max}$ decreases and $T_{\rm max}$ shifts toward the high-temperature side, as observed in the present measurements. We also investigated the effects of $J_2$ and $J_3$, setting the others unity. For $J_2<1$, ${\chi}_{\rm max}$ increases and $T_{\rm max}$ decreases with decreasing $J_2$, whereas for $J_2>1$, the susceptibility diverges for $T{\rightarrow}0$. This is because six spins coupled by $J_2$ on a hexagon form a singlet state and three spins coupled by $J_3$ on a triangle form a doublet state. The susceptibility is not largely affected by $J_3$.

The antiferromagnetic exchange interaction through F$^-$ ion becomes stronger with increasing bonding angle ${\alpha}$ of the exchange pathway Cu$^{2+}$$-$F$^{-}$$-$Cu$^{2+}$. Since ${\alpha}_1$=138.4$^{\circ}$, ${\alpha}_2$=136.4$^{\circ}$, ${\alpha}_3$=133.4$^{\circ}$, and ${\alpha}_4$=123.9$^{\circ}$, the condition $J_1>J_2>J_3>J_4$ must be realized in the present system. Under this condition, we calculated susceptibility. The best fit for $T>T_{\rm max}$ was obtained using $J_1/k_{\rm B}$=234(5)\,K, $J_2/k_{\rm B}$=211(5)\,K, $J_3/k_{\rm B}$=187(5)\,K, and $J_4/k_{\rm B}$=108(5)\,K with $g_{\parallel}$=2.44(1) and $g_{\perp}$=2.15(1). Solid lines in Fig. \ref{fig:SusEx} indicate the susceptibilities calculated with these parameters. These exchange parameters are valid from the fact that $J/k_{\rm B}$=244\,K and ${\alpha}$=141.6$^{\circ}$ in Cs$_2$Cu$_3$ZrF$_{12}$,\cite{Morita} and $J/k_{\rm B}$=103\,K and ${\alpha}$=129.1$^{\circ}$ in KCuGaF$_6$.\cite{Morisaki} For $T<T_{\rm max}$, the calculated susceptibility decreases more rapidly than the experimental susceptibility. This should be ascribed to the finite-size effect. Calculation for a larger cluster may give a better description of low-temperature susceptibility.

\begin{figure}
\begin{center}
\includegraphics[width=8.0cm]{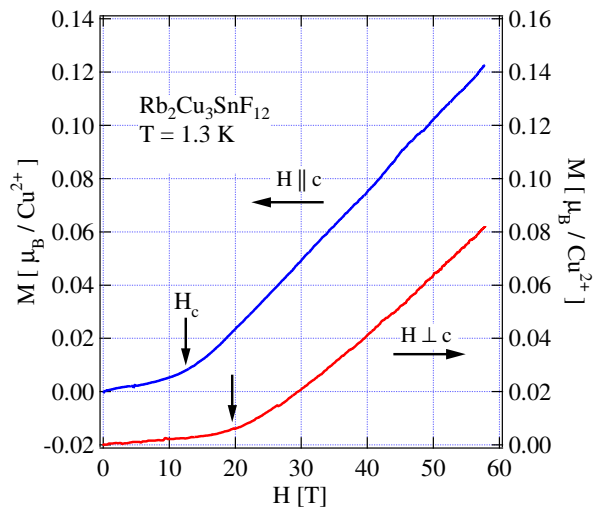}
\end{center}
\caption{Magnetization curve of Rb$_2$Cu$_3$SnF$_{12}$ measured at $T$=1.3 K for $H{\parallel}c$ and $H{\perp}c$. Arrows indicate the critical field $H_{\rm c}$.}
\label{fig:MH}
\end{figure}

To evaluate the spin gap directly, we performed high-field magnetization measurements. Figure \ref{fig:MH} shows magnetization curves measured at $T$=1.3 K for $H{\parallel}c$ and $H{\perp}c$. The magnetization is small up to the critical field $H_{\rm c}$ indicated by arrows and increases rapidly. The levels of the ground and excited states cross at $H_{\rm c}$. The magnetization does not show a sharp bend at $H_{\rm c}$, but is rather rounded. We infer that the antisymmetric interactions, such as the staggered Zeeman and DM interactions, give rise to the smearing of the magnetization anomaly. We assign the critical field $H_{\rm c}$ to the field of inflection in $dM/dH$. The critical fields obtained for $H{\parallel}c$ and $H{\perp}c$ are $H_{\rm c}$=13(1)\,T and 20(1)\,T, respectively. These critical fields do not agree when normalized by the $g$-factor as $(g/2)H_{\rm c}$. When an external field is applied perpendicular to the $c$-axis, the magnetic susceptibility is finite even at $T$=0. Consequently, the ground state energy is not independent of the external field but decreases with the external field, resulting in an increase in the critical field. Therefore, as the spin gap, we take ${\Delta}/k_{\rm B}$=21(1)\,K obtained from $H_{\rm c}$=13(1)\,T for $H{\parallel}c$. This spin gap is the same as that evaluated from low-temperature susceptibility and is approximately one-tenth of $J_{\rm av}$.

Since the largest interaction is the $J_1$ interaction, the ground state is composed of singlet dimers on the $J_1$ bonds. Although the $J_2$ and $J_3$ interactions are smaller than the $J_1$ interaction, their magnitudes are similar. Thus, the local change of dimer positions does not cost so large amount of energy as to create a local triplet state. Such singlet excitations may propagate in a crystal as singlet wave. We infer that there are many singlet states inside the spin gap, as predicted for the proper HKAF, which should be verified by specific heat measurement.

In conclusion, we have presented the results of structural analysis and magnetic measurements of Rb$_2$Cu$_3$SnF$_{12}$. The crystal structure is hexagonal and closely related to the Cs$_2$Cu$_3$ZrF$_{12}$ structure with proper Kagom\'{e} layers of Cu$^{2+}$. Rb$_2$Cu$_3$SnF$_{12}$ can be described as a 2D $S$=1/2 modified HKAF with four kinds of neighboring exchange interaction. The results of magnetic susceptibility and high-field magnetization measurements revealed that the ground state is a disordered singlet with a spin gap, as predicted from a recent theory. We performed exact diagonalization for a 12-site Kagom\'{e} cluster to analyze the magnetic susceptibility and evaluated individual exchange interactions. 

\section*{Acknowledgment}
We express our sincere thank to G. Misguich for showing us his theoretical calculations. This work was supported by a Grant-in-Aid for Scientific Research from the Japan Society for the Promotion of Science, and by a 21st Century COE Program at Tokyo Tech ``Nanometer-Scale Quantum Physics'' and a Grant-in-Aid for Scientific Research on Priority Areas ``High Field Spin Science in 100 T'' both from the Ministry of Education, Culture, Sports, Science and Technology, Japan.

\end{document}